\documentclass[12pt,preprint]{aastex}

\shorttitle{NGS GLAO}
\shortauthors{Baranec et al.}

\begin{document}


\title{Ground-layer wavefront reconstruction from multiple natural guide stars}


\author{Christoph Baranec, Michael Lloyd-Hart and N. Mark Milton}
\affil{Steward Observatory, University of Arizona, Tucson, AZ 85721}
\email{baranec@as.arizona.edu}



\begin{abstract}
Observational tests of ground layer wavefront recovery have been made in open loop using a constellation of four natural guide stars at the 1.55 m Kuiper telescope in Arizona. Such tests explore the effectiveness of wide-field seeing improvement by correction of low-lying atmospheric turbulence with ground-layer adaptive optics (GLAO). The wavefronts from the four stars were measured simultaneously on a Shack-Hartmann wavefront sensor (WFS). The WFS placed a 5 $\times$ 5 array of square subapertures across the pupil of the telescope, allowing for wavefront reconstruction up to the fifth radial Zernike order. We find that the wavefront aberration in each star can be roughly halved by subtracting the average of the wavefronts from the other three stars. Wavefront correction on this basis leads to a reduction in width of the seeing-limited stellar image by up to a factor of 3, with image sharpening effective from the visible to near infrared wavelengths over a field of at least 2 arc minutes. We conclude that GLAO correction will be a valuable tool that can increase resolution and spectrographic throughput across a broad range of seeing-limited observations.
\end{abstract}


\keywords{atmospheric effects --- instrumentation: adaptive optics}



\section{Introduction}

Ground layer adaptive optics (GLAO) is a potentially powerful, but as yet unproven, adaptive optics (AO)
technique which promises modest wavefront correction over wide fields of view. By measuring and averaging
the incoming wavefronts to a telescope at several different field points, an estimate of the common turbulence
located near the entrance pupil of the telescope can be made, with the uncorrelated higher altitude contributions
averaged away. This estimate, when applied to a deformable mirror conjugated near the telescope's pupil,
can correct the atmospheric aberration close to the telescope which is common to all field points.
It has been found empirically at various sites that typically half to two-thirds of the atmospheric
turbulence lies in this ground layer 
(Andersen et al. 2006; Avila et al. 2004; Egner et al. 2006; Lloyd-Hart et al. 2006b, 2005; Tokovinin \& Travouillon 2006; Tokovinin et al. 2005; Velur et al. 2006; Verin et al. 2000),
so when the technique is applied, the natural seeing will improve substantially over a large field. This will
be of particular value to science programs that until now have not found any advantage in AO. Many observations that are normally carried out in the seeing limit will benefit from improved resolution and signal-to-noise, ultimately increasing scientific throughput.

Ground-layer correction was first suggested by Rigaut (2001) as a way to improve wide field imaging for large telescopes. Since then, numerous simulations have shown that GLAO can effectively and consistently improve the atmospheric seeing (Andersen, et al. 2006; Le Louarn \& Hubin 2005; Rigaut 2002; Tokovinin 2004a, 2004b). Plans are underway to implement GLAO at several telescopes around the world with a variety of techniques. The European Southern Observatory will build an AO system with multiple sodium laser guide stars (LGS) for the Very Large Telescope (VLT) that can work in GLAO mode \cite{stu06}. A 1 arc minute square corrected field will be imaged with the instrument MUSE with future plans to correct a 7.5 arcminute square field for the instrument HAWK-I \cite{cas06}. The Southern Astrophysical Research telescope is planning to use a single low altitude Rayleigh LGS to recover the effects of low level turbulence, with a 3 arc minute corrected field fed to downstream science instruments \cite{tok04c}. The Gemini North telescope is currently exploring the feasibility of a GLAO system, with a new deformable secondary mirror (DSM), that will also use sodium LGS \cite{sze06}. The Large Binocular Telescope (LBT) is including a GLAO mode as part of its NIRVANA multi-conjugate adaptive optics system which uses DSMs and up to 16 natural guide stars (NGS) \cite{rag03}. Recently, experiments at the 6.5 m Magellan and MMT telescopes have demonstrated open-loop performance of GLAO correction \cite{ath06, bar06, llo06b, llo05}, and it is currently being implemented at the MMT using a constellation of five Rayleigh LGS. Commissioning runs at the MMT, beginning in late 2006, are expected to demonstrate for the first time the real-world closed-loop performance of this new AO technique \cite{llo06c}.

Driven by considerations of aperture size and sky coverage for AO, future extremely large telescopes (ELT) such as the Giant Magellan Telescope \cite{fab06, joh06, llo06a} and the Thirty Meter Telescope  (Ellerbroek et al. 2005; Ellerbroek et al. 2006; Stoesz et al. 2006) already include multi LGS wavefront sensing in their baseline designs. GLAO, as a natural capability of such systems, is expected to give useful image improvement in the near infrared bands over fields of several arc minutes. Because the performance of GLAO is essentially independent of the size of the aperture, there is no qualitative gain for an ELT over an 8-10 m telescope in resolution. GLAO is nevertheless needed to preserve its quantitative advantage of higher resolution compared to the seeing limit.

Despite the plans described above, closed-loop GLAO has yet to be tested at any telescope. In order to explore its quantitative value, we have carried out open-loop wavefront measurements of a close asterism at the 1.55 m Kuiper telescope on Mt. Bigelow in Arizona. The goal of this experiment was to predict on-sky performance in a closed-loop system, and to explore the behavior of the correction with wavelength and other system parameters.

\section{Experimental Design}


To explore the feasibility of GLAO correction, we designed and built a novel optical system to capture wavefront information from multiple sources in the selected field simultaneously by imaging multiple Shack-Hartmann patterns onto a single CCD. To minimize cost, we utilized off-the-shelf optics. Using the equatorial 1.55 m Kuiper telescope's f/13.5 configuration, the camera had a 2.5 arc minute square field of view with the constraint that stars be separated by a minimum of 30 arc seconds so that their patterns do not overlap. While this is smaller than the field that simulations suggest may benefit from GLAO \cite{and06}, it greatly eases constraints on the optical design of the wavefront sensor (WFS). The pupil was divided by a standard lenslet array into a 5 $\times$ 5 grid of square subapertures, 31 cm on a side when projected back to the primary mirror, of which 20 were illuminated. The final plate scale on the camera is 0.57 arc second per pixel. The detector was a 512 $\times$ 512 pixel Kodak KAF-0261E CCD with 20 $\micron$ pixels which were binned 2 $\times$ 2 on-chip. Since the shortest exposure afforded by the camera's internal shutter was 100 ms, we used an external manually operated photographic shutter to shorten exposures down to 33 ms on the sky, which still gave high enough SNR on the camera for our chosen asterism. The time between exposures was approximately 2 s, much larger than the exposure time, so that successive frames are temporally uncorrelated. Data were taken in sequences of 25 frames with 20 dark frames recorded in between each data set for later background subtraction.

The stars used for the experiment form a close asterism in the constellation Serpens Cauda; a DSS image is shown in Figure 1. The four brightest stars range in V magnitude from 9.4 to 10.6, with separations from the central star between 57 and 75 arc seconds.




\section{Data Analysis}

\subsection{Wavefront reconstruction} \label{wavefrontreconstruction}

Data were taken on the night of 2003 June 17. Figure 2 shows an example of the 512 frames of data recorded at 33 ms exposure time over a 67 minute period. The images show four different Shack-Hartmann spot patterns corresponding to the four brightest stars of the asterism with the same geometry as seen on the sky. 

The spot positions in the Shack-Hartmann patterns were calculated by first resampling the image by block replication onto a grid 10 times finer than the CCD pixels and then finding the local peaks in the convolution of the data with a Gaussian of width 1.3 arc seconds; a parabolic fit to each peak determined the centroid position of the corresponding spot \cite{poy03}. The effect of distortion on the peripheral Shack-Hartmann spot patterns, which can be seen as the non-regular spacing of spots in each pattern in figure 2, was corrected explicitly. The variances in tilt of each row and column of subapertures within each pattern, taken across the full ensemble of data, were measured and fit to linear functions; the fitted variance for each row and column was then scaled to the variance in tilt of subapertures in the central pattern. This correction was necessary to account for misalignment seen in the optical setup. If this systematic error is left uncalibrated, we find that GLAO correction of the measured Zernike modes leaves a residual wavefront error for the most distorted star, number 1, of 515 nm RMS. By applying the distortion correction, the GLAO residual is reduced to 437 nm. The uncertainty in the calibration is small compared to the difference between these results, and therefore also small compared to the residual atmospheric wavefront error.

For each subaperture, the calibrated mean spot position over all 512 frames was subtracted from its instantaneous position in order to remove the effects of static aberrations. The subaperture slopes were then calculated by multiplying the corrected differential spot positions in each axis by the measured plate scale on the optical axis.

Wavefronts from each of the four stars were reconstructed from the 40 subaperture slope measurements by using a synthetic reconstructor matrix derived from a model of the pupil on the Shack-Hartmann lenslet array. The reconstructor matrix creates a vector of coefficients for the first 20 Zernike modes (orders 1 through 5) from the input slope measurements.

\subsection{GLAO performance} \label{glaoperformance}

In this experiment, an estimate of the ground-layer turbulence is calculated as the average of three of the stellar wavefronts. The coefficients for each of the 20 Zernike modes are averaged to give the ground layer estimate which is then subtracted from the fourth star's measured Zernike coefficients to calculate the residual error after ground layer correction. Because of the temporally uncorrelated nature of the data, this correction is done for each individual frame and the effects of servo lag are absent from the GLAO corrections. Figure 3 shows an example of the reconstructed phases for each star from a single frame of data, the ground layer estimate computed from the average of stars 1, 3 and 4, and the residual wavefront of star 2 after ground layer correction.

The same analysis can be applied to each of the four stars. Table 1 shows the RMS wavefront error for each star by Zernike order, along with the residual error after ground-layer correction from the other three stars. The angular separation $\alpha$ between the star and the geometric center of the three other stars used for GLAO correction is also given. Table 1 indicates that up to 45$\%$ of the wavefront error from orders 2-5 can be corrected by GLAO for the central star, 2, enclosed by the three outer stars. In the three other cases where $\alpha$ is large and the star to be corrected is outside of the three beacons used in the GLAO average, GLAO still has traction in correcting the aberrated stellar wavefront, indicating that GLAO correction rolls off smoothly outside of the measurement constellation.

The excessive amount of power seen in the uncorrected tilt modes is due to telescope tracking error and is not indicative of the actual atmospheric tilt. The majority of the tilt power comes from jitter in the east-west direction while there is a constant drift in tilt in the north-south direction. By applying a linear correction to the drift in the north-south direction over the 67 minutes during which data were collected, we can estimate the actual single axis atmospheric tilt to be 471 nm. This gives us a conservative lower estimate of 666 nm in both uncorrected first order modes. This also suggests that approximately 69$\%$ of the atmospheric tilt is corrected for star 2.

The predicted fitting error for our particular geometry of reconstruction is approximately 100 nm RMS. Summing the 2nd through 5th order measured uncorrected wavefront errors with the estimated atmospheric tilt gives 700 nm RMS of total atmospheric wavefront error for each beacon, meaning the fitting error accounts for only $\sim1\%$ of the total measured power. We have also modeled the effects of shot and background noise on our slope measurements. Propagating the error in slopes through our synthetic reconstructor leads to an additional $\leq 1\%$ of total RMS error in each of the stars, depending on their brightness. Because of the poor seeing during the time of observations, the measured wavefront signals were considerably higher in power than our noise sources.

The total strength of the atmospheric turbulence can be calculated based on the RMS values of each of the uncorrected modes (with the exception of tilt) in each of the stars. Using the method developed by Chassat (1992), we calculate a mean Fried length for these observations of $r_{0}$ = 8.0 cm at $\lambda$ = 500 nm. We can also estimate the amount of power in the ground layer turbulence by assuming that the residual power in the GLAO corrected modes is attributable to the uncorrected free atmosphere. In this case we calculate $r_{0}^{FA}$ = 14.5 cm from the GLAO residuals from star 2. This leads to a Fried length of $r_{0}^{GL}$ = 10.6 cm in the ground layer, or 63$\%$ of the total atmospheric turbulence, in line with ground layer estimates at other sites.

As a measure of how well the ground layer estimate of the central star is performing, we can compare this to the optimal linear estimate of star 2's wavefront from the other three stars \cite{llo06b}. We assume a linear relation between the wavefront of the star to be corrected and the wavefronts from the other three guide stars represented by the equation

\begin{equation}
{\hat{\mathbf{a}}_i = \mathbf{T}\,\mathbf{b}_i}
\end{equation}

where, for the \textit{i}th frame in a data sequence, $\hat{\mathbf{a}}_i$ is the vector of Zernike polynomial coefficients characterizing the estimate of star 2's wavefront, $\mathbf{b}_i$ is the vector containing the Zernike coefficients of all the other three stars' reconstructed wavefronts, and $\mathbf{T}$ is the optimal linear reconstructor matrix relating the two. We wish to find a reconstructor $\mathbf{T}$ that minimizes $\left<|\mathbf{a}_i - \hat{\mathbf{a}}_i|^{2}\right>$, the squared norm of the difference between star 2's measured wavefront coefficients $\mathbf{a}_i$ and their estimates, averaged over all the frames. 

To investigate the limit of performance permitted by the data in this least squares sense, we have derived $\mathbf{T}$ by a direct inversion of the data, using singular value decomposition (SVD). This approach does not rely on any a priori model of the atmospheric $C_n^2$ profile or knowledge of the noise characteristics. A matrix $\mathbf{B}$ is constructed from 512 data vectors $\mathbf{b}_i$.  $\mathbf{B}$ is well conditioned and so may be inverted with singular value decomposition to give $\mathbf{B}^+$ with no truncation of the singular values. A similar matrix $\mathbf{A}$ is constructed from the corresponding $\mathbf{a}_i$ vectors. The optimal linear reconstructor is then given by:

\begin{equation}
{\mathbf{T} = \mathbf{A}\,\mathbf{B}^+}
\end{equation}

Applying $\mathbf{T}$ to vectors $\mathbf{b}_i$ drawn from the same data set used to compute it yields the best fit solution $\hat{\mathbf{a}}_i$ and characterizes the noise floor in the data.

A comparison can thus be made between the correction from the simple GLAO average of measured Zernike coefficients from the three field stars and the optimal linear estimate. Table 2 shows the residual errors by Zernike radial order after both types of correction.

The least-squares estimator implicitly accounts for correlations in the stellar wavefronts that arise from turbulence at all altitudes, not just close to the ground. The improvement in performance over the simple average, which when taken in quadrature amounts to 138 nm, is likely largely attributable to high-altitude seeing. However, the linear performance penalty of simple GLAO correction is only 17$\%$ for the particular line of sight to star 2. This result is not as dramatic as the results using this same method and open-loop data with LGS at the MMT where the improvement over GLAO with the least-squares estimator correction ranges from 39$\%$ to 48$\%$ \cite{bar06}. This difference in improvement is due to the difference in telescope diameters; the correcting beacons decorrelate much more quickly as a function of height at the Kuiper telescope and are therefore less able to estimate high-altitude seeing. As we describe below, this modest price paid in axial performance buys a substantially wider corrected field of view than the limit, set by the isoplanatic angle, for full on-axis correction.

\subsection{Predicted image quality} \label{predictedimagequality}

\subsubsection{PSF simulations} \label{psfsimulations}

Simulated point spread functions (PSFs) of both the natural seeing and post-GLAO correction can be created based on the recorded Zernike amplitudes and measured coherence length. For each frame of data we create a random pupil phase map obeying Kolmogorov statistics with inner and outer scales of 2 mm and 100 km. In this way, the effects of uncorrected high order aberration and measurement uncertainty are included in the PSF estimates. Each frame's PSF is generated by calculating the power spectrum of the complex amplitude pupil map with the final long exposure PSF image being the sum of all the individual frame PSFs. Since subsequent frames are temporally uncorrelated, servo lag was absent in these PSF simulations.

For the seeing limited case, we fit and subtract the Zernike modes up to radial order 5 from the random pupil phase map. Because of the excessive tilt caused by telescope tracking error, random tilts drawn from a normal distribution with variance matching our estimate of the actual atmospheric tilt are added back to this phase map; in addition, the 2nd through 5th order modes are replaced with the measured modes from the four stars. For the GLAO corrected images, we replace the lowest 5 orders of Zernike modes from our original random phase map with the Zernike modes measured from the residual of our GLAO corrections. Our simulated GLAO PSFs are therefore based on the measured modes themselves, not their statistics, and include the results of correcting the large telescope tracking error on-sky.

\subsubsection{Calculated FWHM} \label{calculatedfwhm}

Two simulated 17 s J band exposures of star 2 are shown in figure 4, a seeing limited image and an expected GLAO corrected image. In J band, the estimated seeing full-width half-maximum (FWHM) under our seeing conditions is 0.91 arc seconds, and with GLAO this reduces the FWHM by roughly a factor of 3 to 0.27 arc seconds. In J band there are approximately 1.3 $r_0$ lengths across a WFS subaperture diameter; explaining why we are adequately sampling the wavefront with our WFS for good GLAO correction. 

The elongation of the corrected PSF in figure 4 is in the long direction of the three other stars with a calculated ratio of the major and minor axes of 1.25. The ratio of the variances in tilt of the two axes is 1.9 and is not a statistical anomaly but an actual effect of the GLAO correction. In simulation, Andersen et al. (2006) see the same effect of PSF elongation with a $\sim8$ arc min diameter constellation of 3 NGS at the 8 m Gemini telescope, with a mean and maximum PSF axis ratio of 1.10 and 1.14 respectively over the GLAO corrected field. Interestingly, they also found that when using a constellation of 5 equally spaced LGS, the PSF morphology was much more consistent over the GLAO field with a mean PSF axis ratio of only 1.02. This is due to both the symmetry of the correcting LGS constellation of beacons, and the cone-effect decreasing the correleation of high-altitude turbulence sensed.

Figure 5 presents the radially averaged profiles of the simulated PSF exposures seen in figure 4 with the addition of the profiles from the other three GLAO corrected stars. Table 3 and figure 6 present the FWHM values for these exposures, including exposures spanning the visible to the near infrared wavelengths.

GLAO performance is much improved over the native seeing at all wavelengths, with most of the performance gains being made at the near infrared wavelengths as the corrected images are approaching the diffraction limit. Figure 7 shows this same data now by spectral wavelength as a function of $\alpha$. Star 2, only 17 arc seconds from the geometric center of the other three stars, shows the best GLAO correction. Performance then degrades monotonically by distance to the corrected star from the center of its GLAO correcting constellation. Star 1, even at 113 arc sec away, still shows a significant amount of image improvement after GLAO correction. As the GLAO correction is pushed into the longer wavebands, the performance as a function of field position starts to flatten out, as can be seen with the H and K band results.

These simulated PSF results are comparable to the estimated performance of the MMT LGS GLAO system. In seeing of $r_0$ = 14.7 cm, the uncorrected PSF FWHMs in the center of the GLAO field in H, J and K of 0.77, 0.65 and 0.54 arc sec respectively dropped to 0.38, 0.16, and 0.12 arc sec \cite{llo06b}. The experiments at the Kuiper telescope and the MMT have two differences that affect the comparison of the results. First, the Kuiper experiment had fewer $r_0$ lengths across a subaperture diameter, 3.9 at 500 nm, than the MMT with 4.9, so the expectation is that GLAO correction at the Kuiper telescope would be able to extend into the shorter wavelengths. Second, the MMT GLAO results, with a much larger diameter of 6.5 m, as opposed to Kuiper's 1.55 m, did not suffer from diffraction effects at longer wavelengths; with the Kuiper PSF FWHMs approaching the diffraction limit.

\subsubsection{Calculated encircled energy} \label{calculatedencircledenergy}

From these same simulated PSF images, we have calculated two figures of merit describing encircled energy; $\theta_{50}$, the radius within which half of the PSF energy is enclosed, and EE$\%$($0\farcs5$), the percentage of total energy within a canonical 0.5 arc second diameter. These are of particular interest for maximizing the resolution and throughput of spectroscopic science data. Encircled energy as a function of radius is plotted in figure 8 for the J-band exposures presented in figures 4 and 5.

In J band, the seeing limited $\theta_{50}$ = 0.57 arc sec, with 14.4$\%$ of the total PSF energy within 0.5 arc sec, and with GLAO, $\theta_{50}$ reduces down to 0.34 arc sec and EE$\%$($0\farcs5$) increases by over a factor of 2 to 39.2$\%$ for star 2. These figures of merit, for the seeing limit and each GLAO corrected star are presented in table 4 for wavelengths from the visible to the near infrared.

For every GLAO corrected star, $\theta_{50}$ decreases and EE$\%$($0\farcs5$) increases relative to the seeing limit with the largest improvements in image quality for stars closer to the center of the GLAO correcting field and at the longer wavelengths.

We have also computed the metrics, EE$\%$($1\farcs0$), for the seeing limited J, H and K band exposures as 42.7$\%$, 45.5$\%$ and 48.3$\%$. This is interesting to note as these are similar to the EE$\%$($0\farcs5$) values for the GLAO corrected J, H and K values of star 2 being 39.2$\%$, 45.1$\%$ and 49.2$\%$. The improvement in encircled energy afforded by GLAO correction can therefore be thought of as either increasing the resolution of observations by a factor of two and keeping the same throughput or increasing the throughput by a factor of three with a constant resolution. The optimization of resolution and throughput will be an important consideration in the design of scientific instruments and observational programs.

\section{Conclusions}

Our experiment and subsequent analysis suggest that ground-layer AO correction will be a powerful tool for reducing the effects of atmospheric seeing over wide fields. We have shown here that even with only 3 beacons, there is enough averaging of the modes and decorrelation of the upper altitude signal to lead to a good estimate of the ground-layer turbulence. GLAO correction can potentially reduce the stellar PSF FWHM at all visible and near-infrared wavelengths; decreasing it by a factor of 3 in J band. However, the use of NGS causes some elongation of the predicted corrected PSFs.  The anticipated image improvement will also benefit spectroscopy, where in the near-infrared the increase in encircled energy afforded by GLAO correction can either increase the resolution of observations by a factor of two with the same throughput or increase the throughput by a factor of three with a constant resolution. The field over which this image improvement is expected is quite large, with the simulated GLAO correction of star 1, 73 arc sec away from the nearest correcting beacon, showing a reduction in the PSF FWHM by almost a factor of 2 in J band. It has been seen here and in other experiments that GLAO correction can always be expected to improve seeing \cite{llo06b}, but will be particularly powerful when the seeing is moderate to poor, $r_0$ $<$ 15 cm at $\lambda$ = 500.

In comparison, open-loop GLAO experiments with LGS at the MMT show slightly different performance estimates\cite{bar06}. In moderate seeing, GLAO correction is expected to decrease the PSF FWHM by a factor of 4 in H and K bands. However, with a larger subaperture diameter to $r_0$ ratio, GLAO correction at the MMT will not perform as well into the shorter visible wavelengths. We would also expect the PSF morphology to be much more consistent across the GLAO field with the use of LGS.

The first attempt at closing an AO loop around a ground layer estimate will be performed with the MMT LGS AO system in early 2007. There, a constellation of five dynamically refocused Rayleigh lasers are projected onto the sky in a pentagon of two arc minute diameter. The beacon signals will be averaged to get an estimate of Zernike orders 2-8, and when combined with the signal from a tilt camera using an electron multiplying CCD, will be used to drive the MMT's DSM. A near-infrared imager PISCES, with 0.11 arc second pixels and a field of view of 110 arc sec will be used to measure the performance over the field of the laser constellation. Early science programs will focus on deep wide field imaging that will benefit from GLAO, and are slated to begin with establishing realistic performance and sensitivity limits on targets of scientific interest in mid 2007 \cite{llo06c}.

With the demonstrated work here and at the MMT, it is clear that GLAO should be of great interest to current large and future ELTs. As seen in GLAO modeling studies for Gemini \cite{and06}, GLAO has the promise to improve the observed seeing over the raw natural seeing. This can be done either with multiple NGS or LGS beacons, or other novel implementations \cite{mor06}, with advantages to each method. Already the GMT and TMT projects are planning to use multiple LGSs in their designs which can be used to support GLAO observations. All current telescopes with plans to install DSMs, the VLT, Magellan and the LBT, can also follow the lead of the MMT system by installing multiple sodium or Raleigh systems to improve the seeing of their telescopes.





\acknowledgments

This work has been supported by the National Science Foundation under grant AST-0138347. We are grateful for assistance from Johanan Codona, James Georges, Tom Stalcup and Matt Rademacher.

\clearpage

\begin{deluxetable}{crrrrrr}
\tablecolumns{7}
\tablewidth{0pc}
\tablecaption{Wavefront Aberration Before and After GLAO Correction}
\tablehead{
\colhead{}    &  \multicolumn{5}{c}{Zernike Order} & \colhead{$\alpha$} \\
\cline{2-6}  \\
\colhead{Star} & \colhead{1}   & \colhead{2}    & \colhead{3} &
\colhead{4}    & \colhead{5}   & \colhead{(\arcsec)}    }
\startdata
1 & 3468         & 215          & 144          & 93           & 92           & \\
  & \textit{382} & \textit{146} & \textit{109} & \textit{ 75} & \textit{ 78} & 113 \\
2 & 3459         & 226          & 144          & 92           & 91           & \\
  & \textit{206} & \textit{117} & \textit{ 82} & \textit{ 56} & \textit{ 57} &  17 \\
3 & 3467         & 232          & 146          & 92           & 91           & \\
  & \textit{276} & \textit{122} & \textit{ 85} & \textit{ 60} & \textit{ 60} &  65 \\
4 & 3461         & 241          & 157          & 98           & 100          & \\
  & \textit{300} & \textit{142} & \textit{102} & \textit{ 72} & \textit{ 72} &  85 \\
\enddata

\tablecomments{RMS stellar wavefront error in nm, summed in quadrature over all the modes in each Zernike order. Italicized numbers denote residual wavefront error after GLAO correction by the three other stars. The last column reports the angular separation $\alpha$ between the star and the geometric center of the three other stars used for GLAO correction.}

\end{deluxetable}

\clearpage

\begin{deluxetable}{lrrrrrr}
\tablecolumns{7}
\tablewidth{0pc}
\tablecaption{Comparison of AO Correction Methods}
\tablehead{
\colhead{}    &  \multicolumn{6}{c}{Zernike Order} \\
\cline{2-7}  \\
\colhead{Correction Method} & \colhead{1}   & \colhead{2}    & \colhead{3} &
\colhead{4}    & \colhead{5}   & \colhead{1-5}    }
\startdata
No correction                & 3459 & 226 & 144 & \,92 & \,91 & \\
GLAO                         &  206 & 117 &  82 & 56 & 57 & 263\\
Optimal linear reconstructor &  173 & 102 &  70 & 50 & 52 & 224\\
\enddata

\tablecomments{RMS stellar wavefront error in nm by Zernike order for the uncorrected, GLAO corrected and optimally corrected wavefront of star 2.}

\end{deluxetable}

\clearpage

\begin{deluxetable}{lcccccc}
\tablecolumns{7}
\tablewidth{0pc}
\tablecaption{Stellar FWHM for Simulated PSFs}
\tablehead{
\colhead{}    &  \multicolumn{6}{c}{FWHM}  \\
\colhead{}    &  \multicolumn{6}{c}{(\arcsec)}  \\
\cline{2-7}  \\
\colhead{$\lambda$} & \colhead{Seeing Limit}   & \colhead{Diffraction Limit} & \multicolumn{4}{c}{GLAO Corrected Images}   \\
\cline{4-7}  \\
\colhead{} & \colhead{} & \colhead{} & \colhead{Star 1} & \colhead{Star 2} & \colhead{Star 3} & \colhead{Star 4}}

\startdata
500 nm           & 1.14 & 0.07 & 0.82 & 0.66 & 0.72 & 0.79\\
750 nm           & 1.02 & 0.10 & 0.64 & 0.37 & 0.45 & 0.57\\
1 \micron        & 0.95 & 0.13 & 0.54 & 0.28 & 0.34 & 0.43\\
J (1.25 \micron) & 0.91 & 0.17 & 0.48 & 0.27 & 0.32 & 0.38\\
H (1.6 \micron)  & 0.86 & 0.21 & 0.45 & 0.29 & 0.32 & 0.37\\
K (2.2 \micron)  & 0.82 & 0.29 & 0.45 & 0.34 & 0.37 & 0.40\\
\enddata

\end{deluxetable}

\clearpage

\begin{deluxetable}{lccccc}
\tablecolumns{6}
\tablewidth{0pc}
\tablecaption{Stellar Encircled Energy Metrics for Simulated PSFs}
\tablehead{
\colhead{}    &  \multicolumn{5}{c}{$\theta_{50}$ / EE$\%$($0\farcs5$)}  \\
\cline{2-6}  \\
\colhead{$\lambda$} & \colhead{Seeing Limit}  & \multicolumn{4}{c}{GLAO Corrected Images}   \\
\cline{3-6}  \\
\colhead{}    & \colhead{}  & \colhead{Star 1} & \colhead{Star 2} & \colhead{Star 3} & \colhead{Star 4}}
\startdata
500 nm           & 0.69/\,\,\,9.7  & 0.59/14.4 & 0.54/17.7 & 0.55/16.9 & 0.57/15.4 \\
750 nm           & 0.64/11.6 & 0.51/19.5 & 0.45/25.5 & 0.47/24.0 & 0.49/21.4 \\
1 \micron        & 0.60/13.1 & 0.46/24.0 & 0.40/33.2 & 0.41/30.7 & 0.44/27.0 \\
J (1.25 \micron) & 0.57/14.4 & 0.42/27.8 & 0.34/39.2 & 0.36/36.2 & 0.40/31.5 \\
H (1.6 \micron)  & 0.54/15.7 & 0.38/31.7 & 0.29/45.1 & 0.31/41.4 & 0.35/36.3 \\
K (2.2 \micron)  & 0.52/17.3 & 0.35/34.9 & 0.26/49.2 & 0.28/44.7 & 0.31/40.0 \\
\enddata

\tablecomments{First number is $\theta_{50}$ in arc sec and second is encircled energy $\%$ within a 0.5 arc sec diameter.}

\end{deluxetable}



\clearpage

\begin{figure}
\plotone{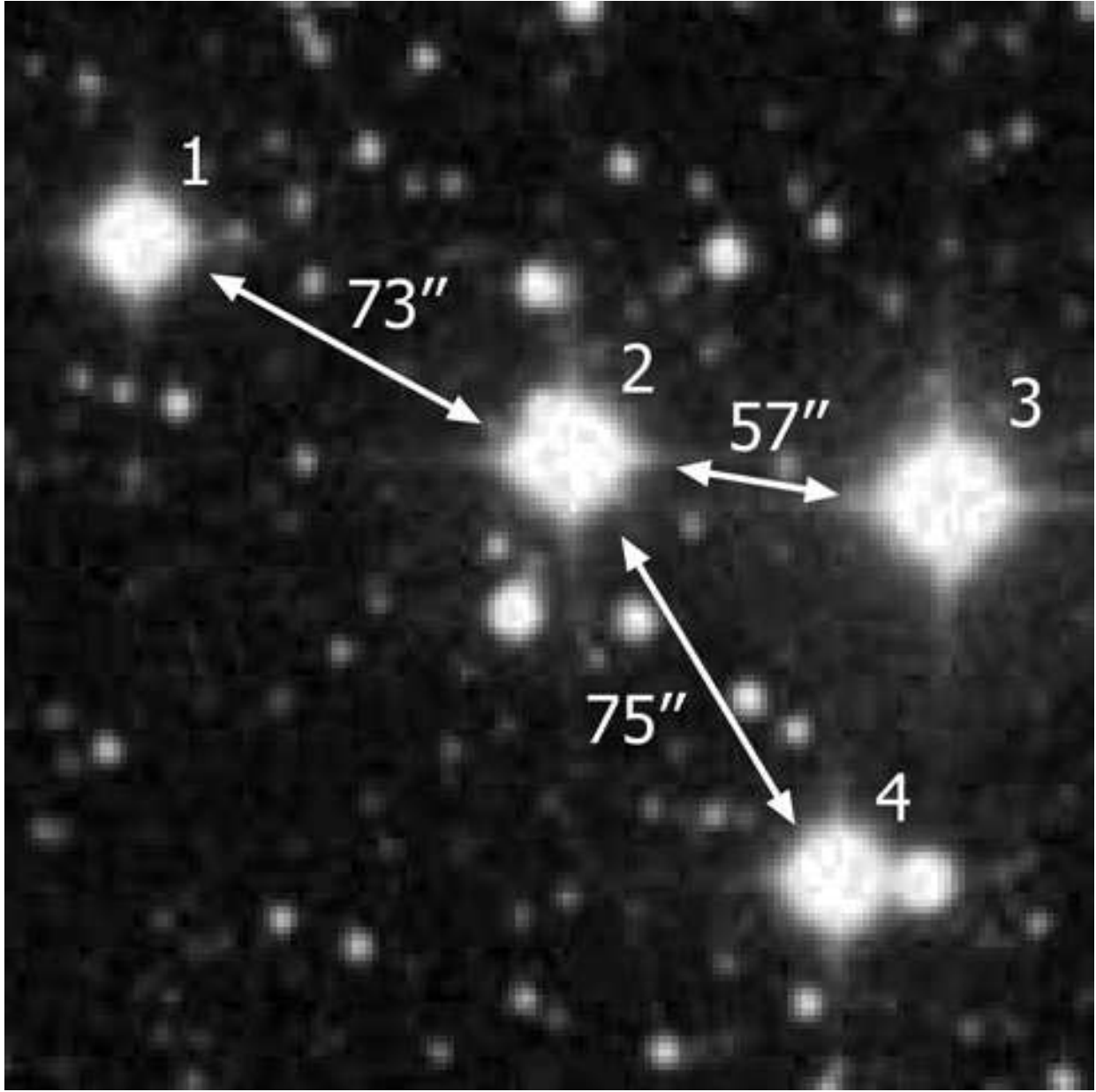}
\caption{DSS2.J.POSSII image of the target asterism.\label{fig1}}
\end{figure}

\clearpage

\begin{figure}
\plotone{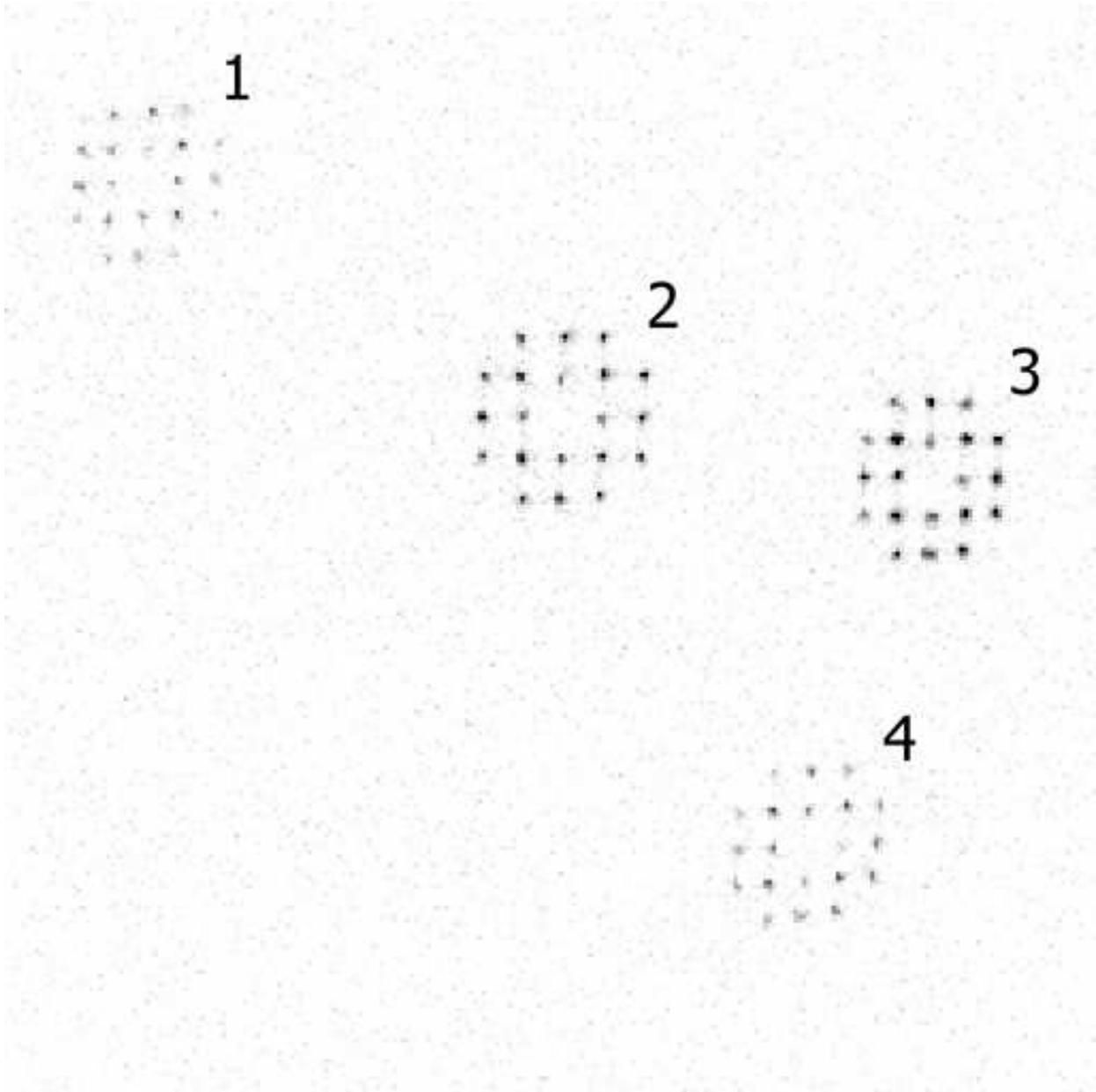}
\caption{Example frame from the Shack-Hartmann wavefront sensor. Each of the four patterns corresponds to one of the stars seen in figure 1.}
\end{figure}

\clearpage

\begin{figure}
\plotone{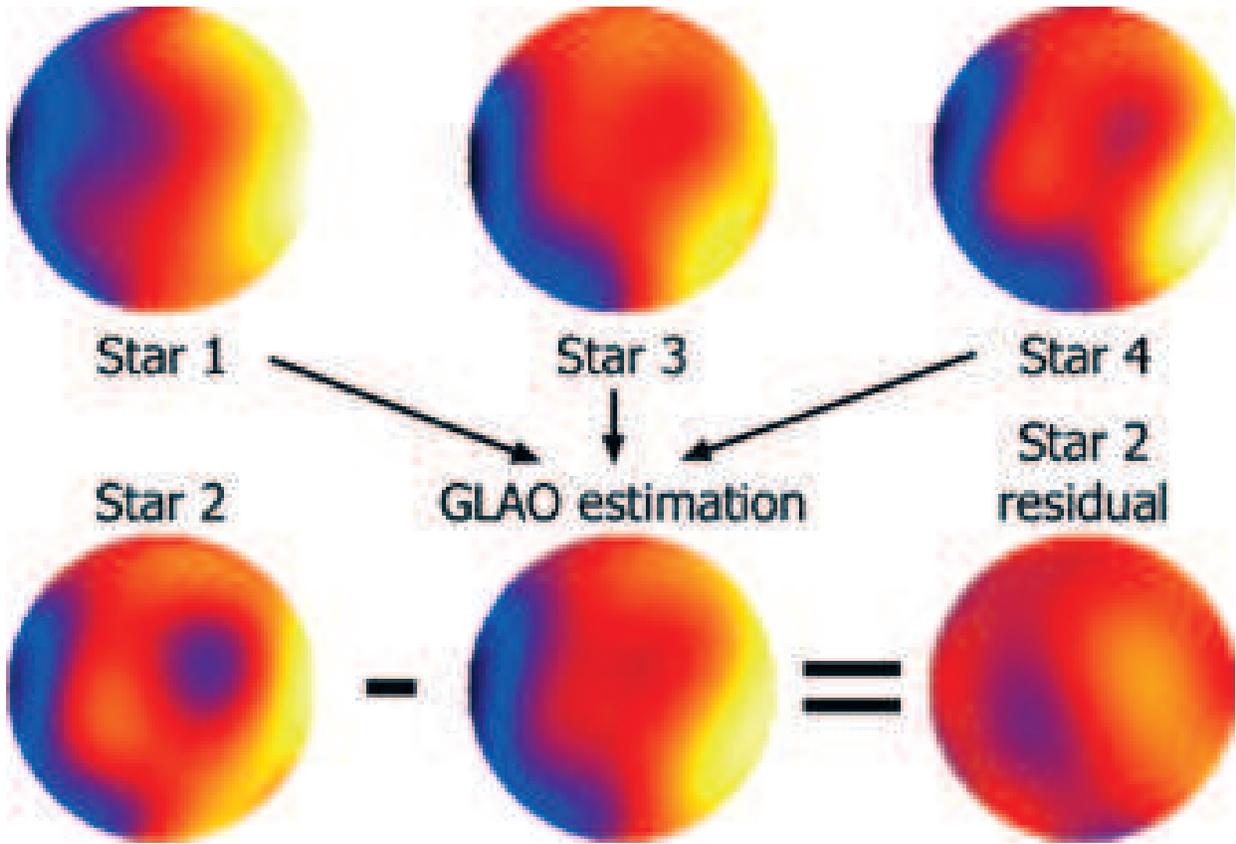}
\caption{Reconstructed phase maps for a single frame for each star, the ground layer estimate based on the average of wavefronts from stars 1, 3 and 4, and the residual of star 2's phase after GLAO correction. Scale is +/- 1.9 $\micron$.}
\end{figure}

\notetoeditor{This figure to appear in color in print.}

\clearpage

\begin{figure}
\plotone{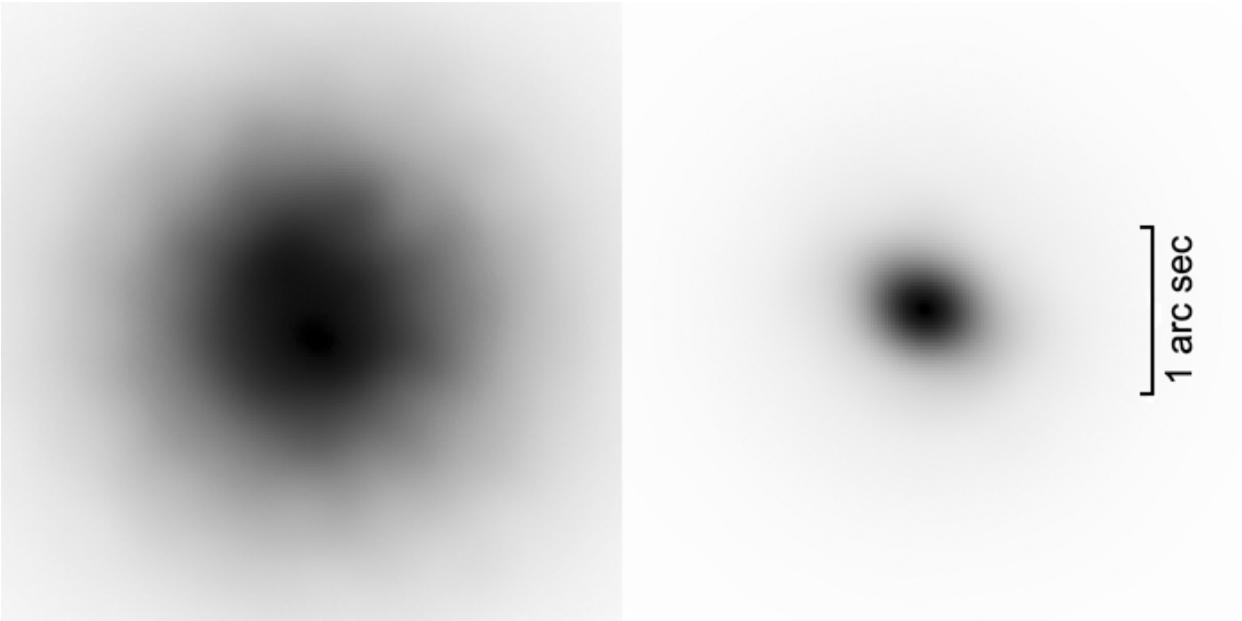}
\caption{Simulated J band PSF of Star 2, with (right) and without (left) GLAO correction. Both PSFs have been scaled by their respective peak intensities; the peak intensity of the GLAO corrected PSF is 6.0 times greater than the peak intensity of the seeing limited PSF. Linear scaling.}
\end{figure}

\clearpage

\begin{figure}
\plotone{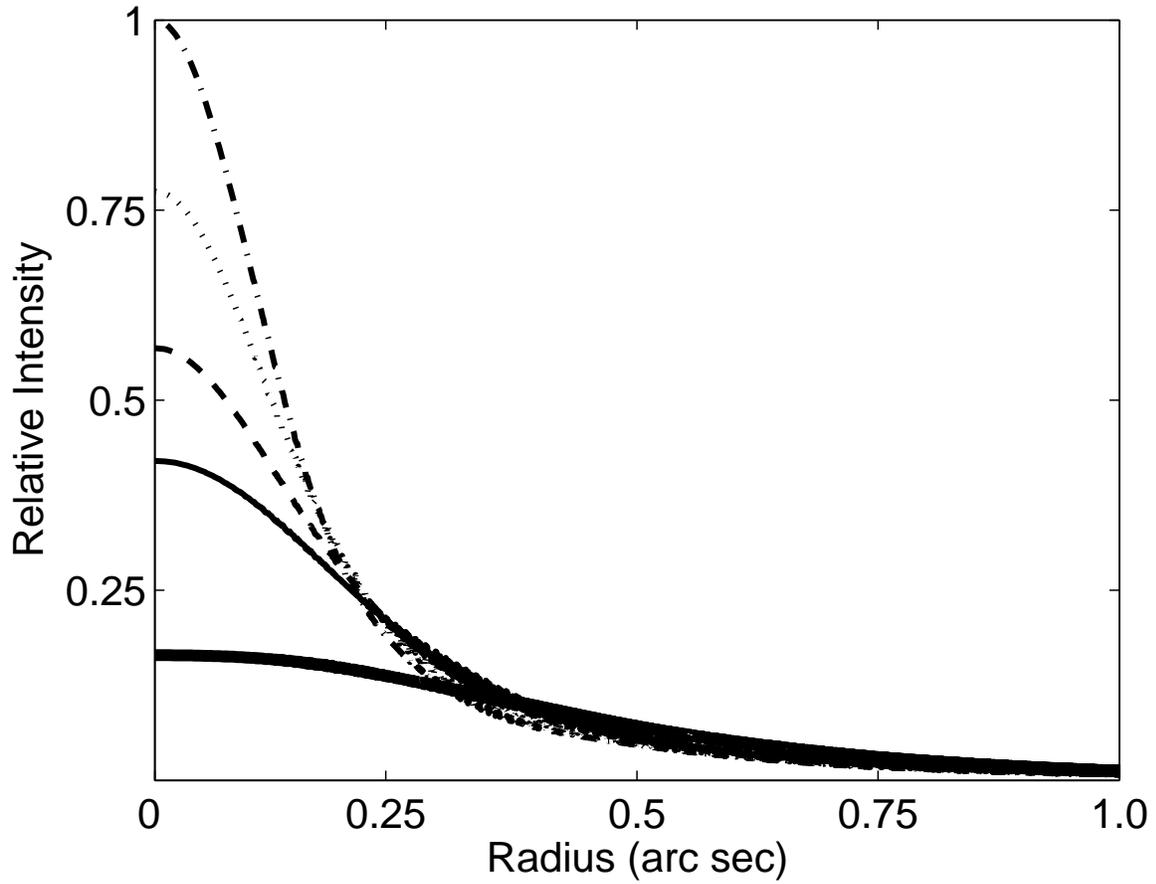}
\caption{Radially averaged J band PSF profiles. Seeing limited PSF in thick solid line, GLAO corrected in thin lines; star 1 (solid), star 2 (dot-dashed), star 3 (dotted), and star 4 (dashed).}
\end{figure}

\clearpage

\begin{figure}
\plotone{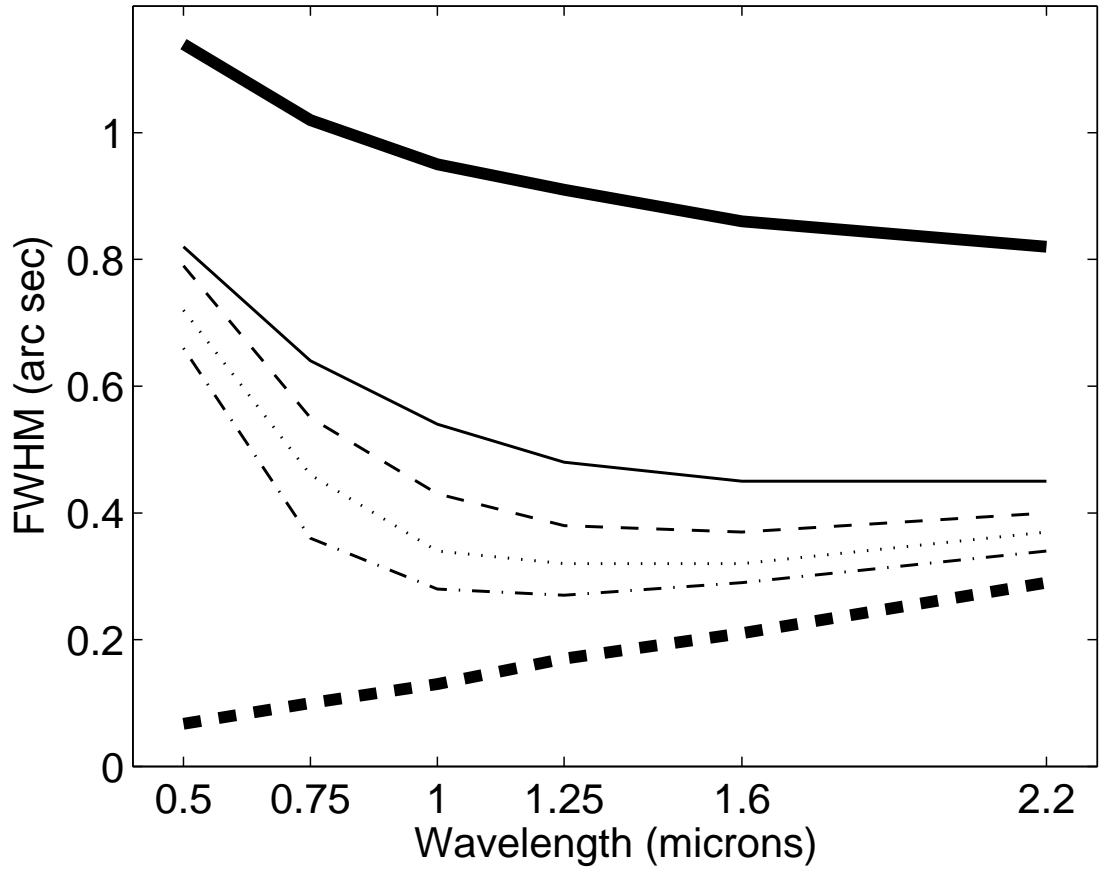}
\caption{FWHM vs. wavelength of simulated PSFs. Seeing limited PSF (thick solid), diffraction limited (thick dashed), and GLAO corrected in thin lines; star 1 (solid), star 2 (dot-dashed), star 3 (dotted), and star 4 (dashed).}
\end{figure}

\clearpage

\begin{figure}
\plotone{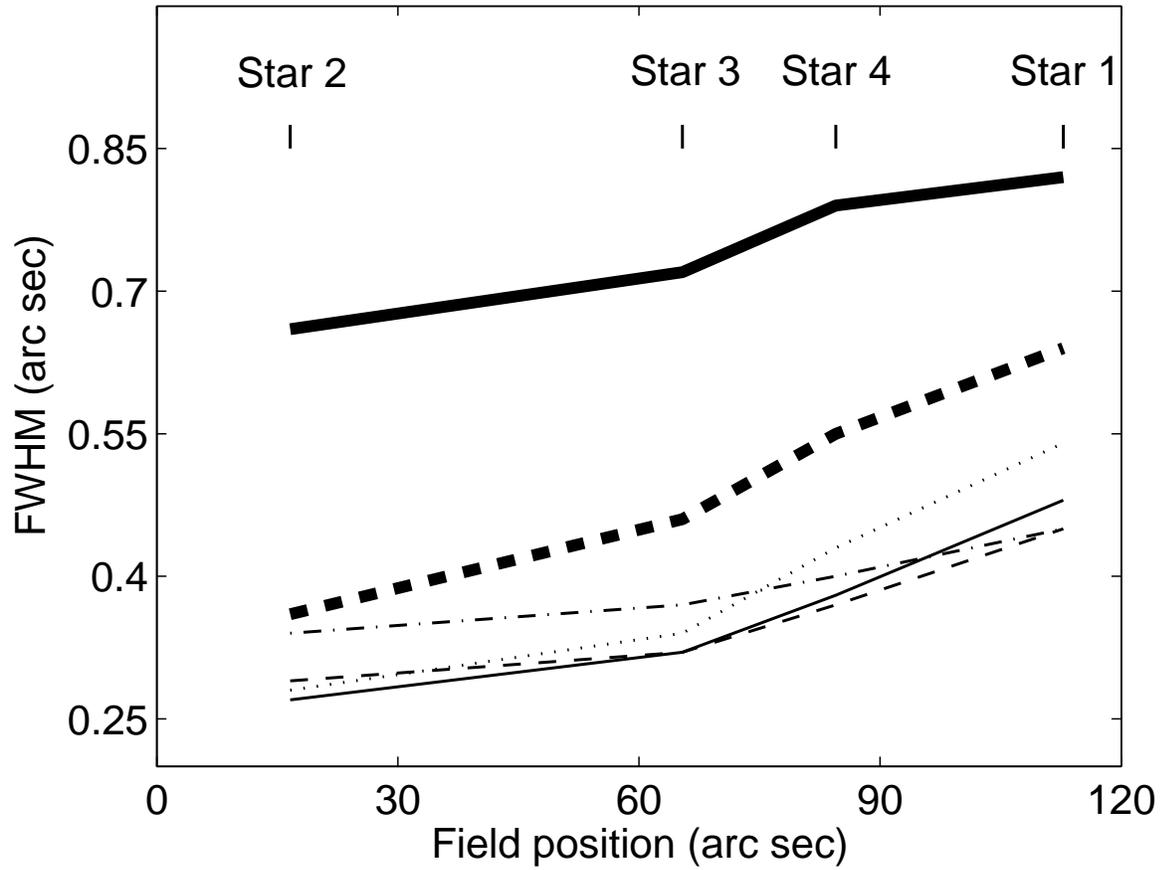}
\caption{Plot of FWHM by wavelength as a function of distance to the center of the GLAO constellation of stars: 500 nm (thick solid), 750 nm (thick dashed), 1 $\micron$ (thin dotted), J (thin solid), H (thin dashed), and K (thin dot-dashed).}
\end{figure}

\clearpage

\begin{figure}
\plotone{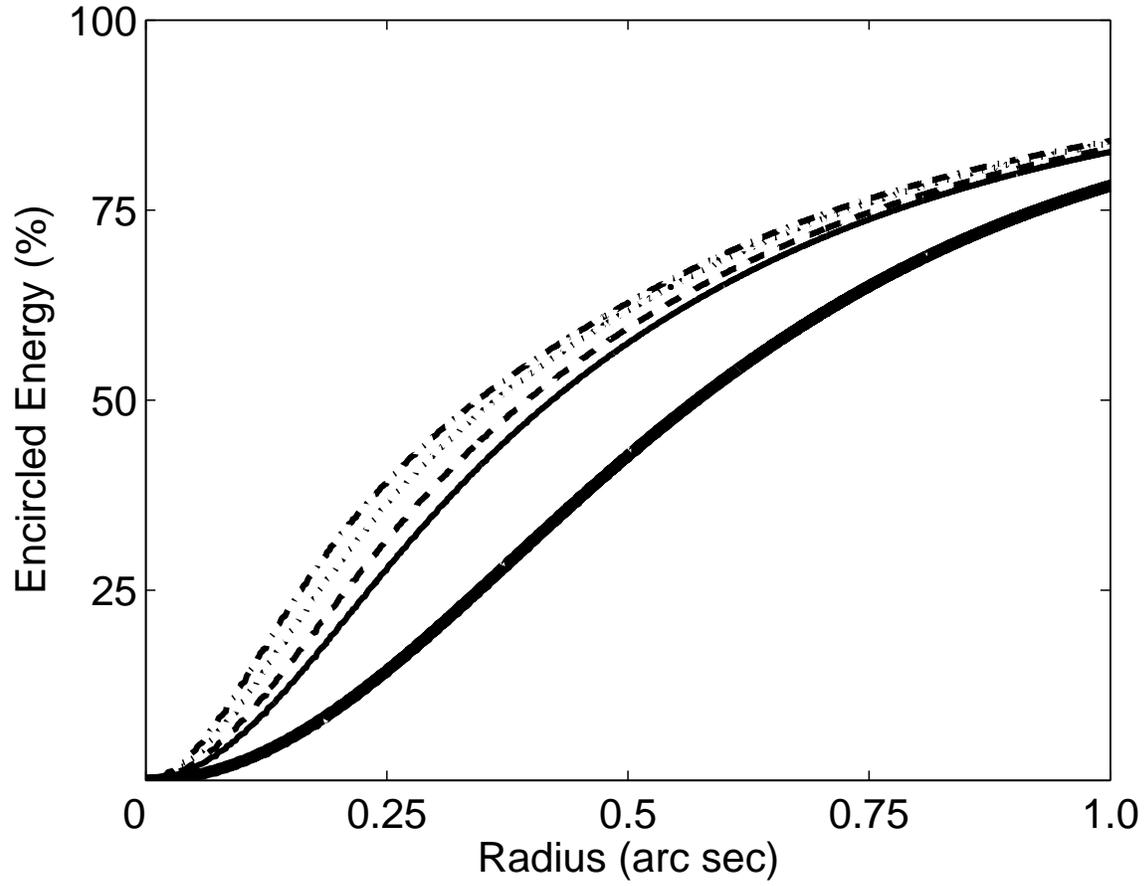}
\caption{PSF $\%$ encircled energy as a function of radius. Seeing limited PSF in thick solid line, GLAO corrected in thin lines; star 1 (solid), star 2 (dot-dashed), star 3 (dotted), and star 4 (dashed).}
\end{figure}

\clearpage

\end{document}